\documentstyle[aps,prb,epsf,amssymb]{revtex}

\begin{document}

\title{Asymptotic temperature dependence of the superfluid density in
liquid $^4$He}

\footnotetext{To be published in Phys. Rev. B\,{\bf 59}, xxxx (1999).}

\author{Torsten Fliessbach\cite{email}}

\address{Fachbereich Physik, University of Siegen, 
 D-57068 Siegen, Germany}

\maketitle
\begin{abstract}

In a modified ideal Bose gas model we derive an expression for the
temperature dependence of the superfluid fraction in liquid $^4$He.
This expression leads to a fit formula for the asymptotic temperature
dependence that reproduces the data significantly better than
comparable formulas.

\bigskip



\end{abstract}

\section{Introduction}
\label{s1}
The behavior of $^3$He and $^4$He liquid on one hand, and of the ideal
Fermi and Bose gas on the other hand, strongly suggests that there is
an intimate relation\cite{fe53} between the $\lambda$ transition in
$^4$He and the Bose-Einstein condensation (BEC) of the ideal Bose gas
(IBG).  Because of the neglect of the interactions one will not expect
that the IBG reproduces all properties of liquid $^4$He, in particular
not those properties that are directly related to the interactions
(like the specific heat or the compressibility). There are, however,
some basic properties of liquid $^4$He that may be
explained\cite{lo54,pu74} by the IBG (like the irrotational superfluid
flow). We start by discussing a discrepancy between the IBG and liquid
$^4$He that is ---as we will point out--- disturbing in view of the
suggested intimate relation between the BEC and the $\lambda$
transition.

The critical behavior of the condensate fraction of the IBG is
\begin{equation}
\label{beta}
\frac{\rho_0}{\rho} \sim |t|^{ 2\beta},\qquad 
\beta = \frac{1}{2}\;,
\end{equation}
where $t = (T - T_\lambda) / T_\lambda$ is the relative temperature;
the IBG transition temperature is equated with that of the $\lambda$
transition. The condensate fraction is commonly identified with the
superfluid fraction; this identification explains\cite{lo54,pu74} a
number of experimental findings of which the most important
one\cite{no90} is that a superfluid current has no vortices. In
contrast to Eq.\,(\ref{beta}), the experimental superfluid fraction
behaves like
\begin{equation}
\label{nu}
\frac{\rho_{\rm s}}{\rho} \sim |t|^{ 2 \nu},\qquad 
\nu\approx \frac{1}{3}\; .
\end{equation}
The suggested intimate connection between the BEC and the
$\lambda$ transition is in conflict with $\beta \ne \nu$. The values
$\beta=1/2$ and $\nu\approx 1/3$ imply $\rho_0 \ll \rho_{\rm s}$ just
below the transition.

The standard solution of the conflict between Eqs.\,(\ref{beta}) and
(\ref{nu}) appears to be the renormalization-group method. In this
approach one starts from a Ginzburg-Landau ansatz for the free energy
(or enthalpy) that leads to the critical exponent $1/2$ for the order
parameter. For the considered universality class the renormalization
procedure yields then values near to $1/3$ for the critical exponent of
the order parameter. This may well serve as an explanation of the
experimental value of $\nu\approx 1/3$ in Eq.\,(\ref{nu}) but it does
not resolve the conflict between Eq.\,(\ref{beta}) and Eq.\,(\ref{nu}):
A renormalization is appropriate for the Landau value $\beta =1/2$ but
not for the IBG value $\beta=1/2$. The reason is that the IBG value is
obtained by an exact evaluation of the partition sum.  The exact
evaluation of the partition sum implies a summation over arbitrarily
small momenta (or, correspondingly, arbitrarily large distances).
Therefore, the reasoning behind the renormalization procedure (analytic
Ginzburg-Landau ansatz for {\em finite}\/ regions or blocks, and
subsequent transformation to larger and larger blocks) cannot be
applied to the IBG free energy.

Moreover, the critical exponent $\beta =1/2$ of the IBG cannot be
changed within the IBG frame without destroying the mechanism leading
to the BEC. The exponent $\beta =1/2$ is characteristic for the IBG
with the BEC phase transition.

We have now discussed two points: (i) We expect an intimate relation
between the Bose-Einstein condensation and the $\lambda$ transition.
(ii) The IBG value $\beta=1/2$ should be taken seriously (because it is
a result of an exact evaluation of a partition sum, and because
$\beta\ne 1/2$ is not compatible with the BEC mechanism). From these
two points we conclude the following: The theoretical $\beta=1/2$ in
Eq.\,(\ref{beta}) and the experimental $\nu \approx 1/3$ in
Eq.\,(\ref{nu}) are in conflict with each other.

Within the frame of the ideal Bose gas model we propose to resolve this
conflict by the assumption that {\em noncondensed particles move
coherently with the condensate}\/. This means that we no longer
identify the condensate with the superfluid fraction; the condensate is
only part of the superfluid phase. A coherent motion can be described
by multiplying the real single particle functions of noncondensed
particles by the complex phase factor of the condensate. The superfluid
density $\rho_{\rm s}$ is then made up by the condensate density
$\rho_0$ plus the density $\rho_{\rm coh}$ of the coherently comoving,
low momentum noncondensed particles. This concept leads to an
expression and eventually to a fit formula for the temperature
dependence of the superfluid density.

We will stick to the essential characteristic  of the IBG (in
particular the BEC) but introduce some modifications (for example,
Jastrow factors) that are necessary for a realistic approach to liquid
$^4$He. This modified IBG is called {\em almost ideal Bose gas model}\/
(AIBG). The AIBG has been introduced some years ago as an attempt to
explain the (nearly) logarithmic singularity of the specific
heat\cite{fl91}. Some consequences of the decomposition $\rho_{\rm s} =
\rho_0 + \rho_{\rm coh}$ have been discussed in Refs. \onlinecite{fl91}
and \onlinecite{sc94}.  The present paper is devoted to the
investigation of the temperature dependence of the superfluid density
in this model. The necessary details of the underlying model, the AIBG,
will be given below.

The form of the temperature dependence of the superfluid fraction is
derived in Sec.\,\ref{s2}. This leads to a fit formula for the
temperature dependence of $\rho_s$ that is applied to experimental data
and compared to other fit formulas (Sec.\,\ref{s3}).  Sec.\,\ref{s4}
discusses the temperature dependence of the condensate density.
Section \ref{s5} presents scaling arguments on the basis of an
effective Ginzburg-Landau model; this includes a qualitative
explanation of the coherent comotion of noncondensed particles and
leads to restrictions for some of the parameters of the fit formula.

\section{AIBG form of the superfluid fraction}
\label{s2}

\subsection{Many-body wave function}
\label{s2.1}
Following Chester\cite{ch55} we multiply the IBG wave function
$\Psi_{\rm IBG}$ by Jastrow factors $F = \prod f_{ij}$,
\begin{equation}
\label{chester}
\Psi =  F\,\Psi _{\rm IBG}=  \prod_{i<j}^N f_{ij}(r_{ij}) \, 
\Psi _{\rm IBG} ({\bf r}_1,\ldots,{\bf r}_N; n_{\bf k})
\; .
\end{equation}
We consider $i=1,2,...,N$ atoms in a volume $V$. The occupation numbers
$n_{\bf k}$ are parameters of the wave function; in physical quantities
they are eventually replaced by their statistical expectation values
$\langle n_{\bf k} \rangle$. The Jastrow factors  take into account the
most important effects of the realistic interactions; with a suitable
choice for the $f_{ij}$ (for example, $f_{ij}(r) = \exp [-(a/r)^b]$
with $a$ and $b$ determined by a variational principle\cite{mc65}) the
wave function (\ref{chester}) leads to a realistic pair-correlation
function.

The IBG wave function $\Psi_{\rm IBG}$ in Eq.\,(\ref{chester}) is the
symmetrized product of single-particle functions. We display this
structure admitting at the same time a phase field $\Phi$ of the
condensate:
\begin{equation}
\label{Psi}
\Psi = {\cal S}\,F\,\big[ \exp(\,{\rm i}\,\Phi ) \big]^{n_0}
\prod_{ {\bf k}\,\ne \,0} \left[\, \varphi_{\bf k} \,\right]^{n_{\bf
k}}  \; .
\end{equation}
Here ${\cal S}$ denotes the symmetrization operator. The $\varphi_{\bf
k}$ are the real single-particle functions of the noncondensed
particles. The schematic notation $[\varphi_{\bf k} ]^{n_k}$ stands for
the product $\varphi_{\bf k}({\bf r}_{\nu+1}) \cdot\varphi_{\bf k}({\bf
r}_{\nu + 2})\cdot \ldots\cdot \varphi_{\bf k}({\bf r}_{\nu+n_k}) $;
this notation applies also to $[\exp(\,{\rm i}\,\Phi )]^{n_0}$. All
$n_0$ condensed particles adopt the same phase factor $ \exp(\,{\rm
i}\,
\Phi ({\bf r}))$ forming the macroscopic wave function
\begin{equation}
\label{psi}
\psi ({\bf r}) =\sqrt{ \frac{n_0}{V}\,}\,
 \exp\big[\,{\rm i}\,\Phi ({\bf r})\big]\; .
\end{equation}
The phase field $\Phi$ describes the coherent motion of the condensate
particles. (Actually, one has to construct a suitable coherent
state\cite{an66}. This point is, however, not essential for the
following discussion.) This motion is superfluid if the velocity ${\bf
u}_{\rm s} = \hbar\, \nabla \Phi /m$ is sufficiently small.

Equations (\ref{Psi}) and (\ref{psi}) are a well-known
description\cite{lo54} for a superfluid motion in the IBG. In this
description the superfluid fraction $\rho_{\rm s}/ \rho$ equals the
condensate fraction $n_0/N = \rho_0 / \rho$. The role of the Jastrow
factors in this context will be discussed in Sec.\,\ref{s2.2}.

In order to dissolve the discrepancy between Eqs.\,(\ref{beta}) and
(\ref{nu}) we assume that {\em noncondensed particles move coherently
with the condensate}\/.  This is possible if noncondensed particles
adopt the macroscopic phase of the condensate:
\begin{equation}
\label{Psic}
\Psi = {\cal S}\,F\,\big[ \exp(\,{\rm i}\,\Phi )\big]^{n_0}
\prod_{0\, < \, k\,\le \,k_{\rm coh}} 
\big[\,\varphi_{\bf k}\,\exp(\,{\rm i}\, \Phi ) \,\big]^{n_k} 
\prod_{k\,> \,k_{\rm coh}} \big[\,\varphi_{\bf k} \,\big]^{n_k} \; .
\end{equation}
We assume the phase ordering for all states with momenta below a
certain coherence limit $k_{\rm coh}$. For the low lying states with
$n_k\gg 1$ such phase ordering is relatively easy because it requires
only a small entropy decrease. At this stage, $k_{\rm coh}$ should be
considered as a model parameter. In Sec.\,\ref{s5.1} the existence and
the size of this coherence limit will be made plausible.

We evaluate particle current for the wave function (\ref{Psic}):
\begin{equation}
\label{current}
{\bf j}_{\rm s}({\bf r},n_k) = \left\langle \Psi\, \left\vert
\sum_{n=1}^N\, \hat{\bf j}_n + {\rm c.c.} \,\right\vert \,\Psi
\right\rangle = \frac{\rho}{N}\,\frac{\hbar}{m} \left( n_0 +
{\sum_{k<k_{\rm coh}}}^{\!\!\!\prime}   n_k \right) \nabla \Phi \; .
\end{equation}
The prime at the sum over the momenta $k<k_{\rm coh}$ means that the
$k=0$ contribution is excluded. In coordinate space, the current
operator reads ${\bf j}_n =-{\rm i}\, \hbar\,\nabla_{\! n} /(2\,m) +
{\rm c.c}$. It acts on all ${\bf r}_i$-dependences. Because of the
added conjugate complex term all contributions from the real functions
(the $f_{ij}$ in $F$ or the $\varphi_k$) cancel. The only surviving
terms are those where ${\bf j}_n$ acts on the phase $\Phi$.

For a superfluid motion with ${\bf u}_{\rm s}=\hbar\,\nabla\Phi/m$ and
in the statistical average, ${\bf j}_{\rm s}$ of Eq.\,(\ref{current})
equals $\rho_{\rm s}\, {\bf u}_{\rm s}$. We may then read off the
superfluid fraction,
\begin{equation}
\label{rhos}
\frac{\rho_s}{\rho} = \frac{1}{N}\,\left( \langle n_0\rangle +
{\sum_{k<k_{\rm coh}}}^{\!\!\!\prime} \langle n_k\rangle
\right) = \frac{\rho_0+\rho_{\rm coh}}{\rho} \; .
\end{equation}
This expression will be evaluated in Sec.\,\ref{s2.3}.

The ansatz (\ref{Psic}) leading to Eq.\,(\ref{rhos}) shows in which way
noncondensed particles may contribute to the superfluid density.

\subsection{Condensate density}
\label{s2.2}
We discuss in some detail what is meant by the terminus ``condensate
density'', in particular with respect to the Jastrow factors in
Eqs.\,(\ref{Psi}) or (\ref{Psic}).

The {\em exact condensate density}\/ may be defined by
\begin{equation}
\label{9}
\langle\Psi | \, \hat \phi ^+({\bf r})\,\hat \phi ({\bf r}')\,|\Psi
\rangle 
\stackrel{ |{\bf r}-{\bf r}'| \to \infty}{\longrightarrow}
\rho_0^{\rm exact} \; ,
\end{equation}
where $\Psi$ is the exact many-body state and the $\hat \phi^+$ and
$\hat \phi$ are single-particle creation and annihilation operators.
For finite temperatures one has to take the statistical expectation
value of $\rho_0^{\rm exact}$ (we do not introduce a different symbol).

For an IBG wave function $\Psi_{\rm IBG}$ the condensate density is
given by $\rho_0^{\rm model} = n_0/V$. In the statistical average this
this {\em model condensate density} becomes
\begin{equation}
\label{10}
\rho_0^{\rm model} = \frac{\langle n_0\rangle}{V} 
\; .
\end{equation}

The exact many-body state in Eq.\,(\ref{9}) may be approximated by
Eq.\,(\ref{chester}), or by $\Psi\approx F$ for the ground state. In
this case the relation between both condensate densities is well known:
The model condensate fraction is {\em depleted}\/\cite{pe56,wi87} by
the Jastrow factors $F$, for example, from $\rho_0^{\rm model}/\rho =1$
to $\rho_0^{\rm exact} /\rho\approx 0.1$ for $T=0$.

The above calculation leading to Eq.\,(\ref{rhos}) demonstrates the
following point: In contrast to the density $\rho_0^{\rm model}$, {\em
the current density $\rho_0^{\rm model}\,{\bf u}_{\rm s}$ is not
depleted}\/. The reason is that in Eq.\,(\ref{current}) all derivatives
of the real Jastrow factors cancel (because of the added conjugate
complex term).

On the basis of this point we arrive at the following statements about
the role of the densities $\rho_0^{\rm model}$, $\rho_0^{\rm exact}$,
and $\rho_{\rm s}$.

\begin{enumerate}
\item[(a)]
Since the current density $\rho_0^{\rm model}\,{\bf u}_{\rm s}$ is not
depleted we may identify $\rho_0^{\rm model}$ (and not $\rho_0^{\rm
exact}$) with the square $|\psi|^2$ of the macroscopic wave function.
Irrespective of the Jastrow factors we may use Eq.\ (\ref{psi}) as it
stands. For a superfluid flow, the phase $\Phi({\bf r})$ of the
macroscopic wave function (\ref{psi}) fixes the velocity field ${\bf
u}_{\rm s} =\hbar\, \nabla \Phi /m$. The basic relations for the
superfluidity (like ${\rm curl}\, {\bf u}_{\rm s} =0$ and the
Feynman-Onsager quantization rule) are not affected by the Jastrow
factors in the many-body wave function.

\item[(b)]
The exact condensate density is a quantity of its own right. It is the
density of the zero momentum particles in the liquid helium. Recently
Wyatt\cite{wy98} reported about a rather clear experimental evidence
for this condensate. For a review about the attempts to determine
$\rho_0^{\rm exact}$ experimentally we refer to Sokol\cite{so95}.

\item[(c)]
The assumption that noncondensed particles move coherently with the
condensate is introduced by the step from Eq.\,(\ref{Psi}) to
Eq.\,(\ref{Psic}).  Again, this step does not alter the basic relations
following from Eq.\,(\ref{psi}) (like ${\bf u}_{\rm s} = \hbar\, \nabla
\Phi /m$, ${\rm curl}\, {\bf u}_{\rm s} =0$ and the Feynman-Onsager
quantization rule).

\item[(d)]
For $T=0$ the value $\rho_0^{\rm model}/\rho =1$ yields $\rho_{\rm
s}/\rho=1$ [as we will see, $\rho_{\rm coh}$ in Eq.\,(\ref{rhos})
contributes only in the vicinity of $T_\lambda$]. In contrast to this
the connection between $\rho_0^{\rm exact}/\rho \approx 0.1$ with
$\rho_{\rm s}/\rho=1$ is less obvious. For $T\approx 0$ the value
$\rho_0^{\rm model}/\rho\approx 1$ implies $\rho_{\rm s}/\rho\approx
1$. For describing $1- \rho_{\rm s}/\rho$ quantitatively one must
however include phonons. This is not done in Eqs.\,(\ref{Psi}) or
(\ref{Psic}) because our primary object is the asymptotic temperature
region.

\end{enumerate}

We summarize this subsection: As far as the superfluid current is
concerned the model condensate density is not depleted. The model
condensate density is the fundamental constituent of the superfluid
density.

In the following the model condensate density $\rho_0^{\rm model}$ will
again be denoted by $\rho_0$ and called condensate density.

\subsection{Superfluid density}
\label{s2.3}
We evaluate the expression (\ref{rhos}) for the superfluid density.
Our model assumes expectation values $\langle n_k\rangle$ that are of
the IBG form,
\begin{equation}
\label{11}
\langle n_k\rangle =
\frac{1}{\exp \,[(\,\epsilon_k-\mu )/k_{\rm B}T] -1} \, = \,
\frac{1}{\,\exp \,(x^2+\tau^2)-1}
\; .
\end{equation}
Here $\mu$ is the chemical potential, $\epsilon_k=\hbar^2k^2/2m$ are
the single-particle energies, and $k_{\rm B}$ is Boltzmann's constant.
In the last expression we introduced the dimensionless quantities
$\tau^2=-\mu /k_{\rm B}T$ and
\begin{equation}
\label{12}
x=\frac{\lambda \,|{\bf k}|}{\sqrt{4\pi }}\,,
\quad\mbox{with}\quad
\lambda =\frac{2\pi\hbar}{\sqrt{2\pi m\,k_{\rm B}T}}
\; .
\end{equation}
The transition temperature of the IBG is given by the following
condition for the thermal wave length $\lambda$:
\begin{equation}
\label{13}
\lambda (T_\lambda ) = \big[ \,v \,\zeta (3/2)\,\big]^{1/3}
\; ,
\end{equation}
where $\zeta (3/2) =2.6124$ denotes Riemann's zeta function. In
applying our almost ideal Bose gas model (AIBG) to the real system we
identify $T_\lambda$ with the actual transition temperature. In the
following we use the relative temperature
\begin{equation}
\label{14}
t=\frac{T-T_\lambda }{T_\lambda }
\; .
\end{equation}
We evaluate the condensate density:
\begin{equation}
\label{rho0}
\frac{\rho_0}{\rho} = 
1- {\sum}^{\prime} \; \frac{\langle n_{\bf k}\rangle}{N} = 1 -
(1+t)^{3/2}\; \frac{g_{3/2}(\tau)}{\zeta(3/2)}
\; .
\end{equation}
Riemann's generalized zeta function is given by $g_p(\tau) =
\sum_1^\infty \exp (-n\,\tau^2)/n^p$, and $\zeta(p)=g_p(0)$. 
The chemical potential $\mu$ or, equivalently, $\tau$ may be expanded
for $|t|\ll 1$:
\begin{equation}
\label{16}
\tau (t) =\sqrt{\frac{-\mu}{k_{\rm B}T}} = 
\left\{ \begin{array}{lll}
a\, t+b \,t^2 +\ldots &\quad & (t>0) \\[0.5mm] a'\,|t| + b'\,t^2+\ldots
&&(t<0)
\end{array} \right.
\; .
\end{equation}
For $t>0$ Eq.\,(\ref{rho0}) with $\rho_0/\rho = 0$ yields
$(1+t)^{3/2}\, g_{3/2}(\tau) = \zeta(3/2)$. This condition determines
the temperature dependence of $\tau (t)$ and in particular the
coefficients $a$, $b$, \ldots, for example $a= 3\,\zeta(3/2)/(4
\pi^{1/2})$.

For $t<0$ the IBG yields $\tau=0$. In the AIBG we admit nonvanishing
coefficients $a'$, $b'$,\ldots{} in Eq.\,(\ref{16}). This makes the
expansion (\ref{16}) more symmetric; it corresponds to a
phenonemological gap between the condensate level and the noncondensed
particles. A coefficient $a'\ne 0$ does not affect the BEC as the most
important feature of IBG. It avoids, however, the divergence of the
static structure factor\cite{bl92} $S(k)$ for $k\to 0$  and greatly
improves the unrealistic ($\propto T^{3/2}$) behavior of the specific
heat. In view of the successful roton picture it is not too surprising
that a gap is necessary for a quantitative description of the
superfluid density (or of the specific heat). As we will see, a
realistic description of liquid helium requires $a'\approx 3$; the next
coefficient $b'$ will not be needed.

From Eq.\,(\ref{rho0}) and with Eq.\,(\ref{16}) we obtain
\begin{equation}
\label{17}
\frac{\rho_0}{\rho} = f\,|t| + g\,t^2 +\ldots  \qquad (t<0)
\end{equation}
where
\begin{equation}
\label{18}
f = \frac{3}{2} + \frac{2\sqrt{\pi }\,a'}{\zeta (3/2)}
\; .
\end{equation}

We evaluate now the density of the comoving particles
\begin{equation}
\label{rhoc}
\frac{\rho_{\rm coh}}{\rho}= 
{\sum_{k<k_{\rm coh}}}^{\!\!\!\prime} \; \frac{\langle n_k\rangle}{N} =
\frac{4 \,(1+t)^{3/2}}{\sqrt{\pi}\, \zeta(3/2)}
\int^{x_{\rm coh}}_0 \! \frac{x^2\, dx}{\exp(x^2 +\tau^2)-1}
\; .
\end{equation}
Using $y/[\exp(y)-1] = 1 -y/2  +y^2/12 \mp \ldots$ we obtain
\begin{equation}
\label{20}
\frac{\rho_{\rm coh}}{\rho}
 = \frac{4\,(1+t)^{3/2}}{\sqrt{\pi}\, \zeta(3/2)} \bigg( x_{\rm coh}
-\tau \,\arctan \frac{x_{\rm coh}}{\tau} -
\frac{x_{\rm coh}^{\, 3}}{6} +  \frac{x_{\rm coh}^{\, 5}}{60} +
\frac{x_{\rm coh}^{\, 3}\,\tau^2}{36} 
\pm \ldots \bigg)
\; .
\end{equation}
The convergence of this expression is excellent; for the actual
parameter values and for $|t|\le 0.1$ the terms not shown are of the
order $10^{-8}$.

We have not yet specified the coherence limit $k_{\rm coh}$.
For $|t|\ll 1$ we will find $\rho_{\rm s} \sim \rho_{\rm coh} \sim
k_{\rm coh}$ for the superfluid density and $\rho_{\rm s}\, k_{\rm
coh}^{\,2} \sim k_{\rm coh}^{\, 3}$ for the kinetic energy of the
fluctuations. Requiring that this kinetic energy scales with the free
energy $F\sim - \langle n_0\rangle^2 \sim - t^2$ yields
\begin{equation}
\label{21}
k_{\rm coh}\sim |t|^{2/3}
\; .
\end{equation}
This scaling argument will be presented in more detail in Sec.\,\ref{s5}.

Inserting Eq.\,(\ref{21}) in Eq.\,(\ref{20}) and using Eq.\,(\ref{17}),
the superfluid fraction contains the powers $|t|^{2/3}$, $|t|$,
$|t|^{4/3}$, and so on:
\begin{equation}
\label{22}
\frac{\rho_s}{\rho} =
\frac{\rho_0 + \rho_{\rm coh}}{\rho} = 
 a_1\,|t|^{2/3} +a_2\,|t| +a_3\,|t|^{4/3} 
+ \ldots
\; .
\end{equation}

\subsection{AIBG assumptions}
\label{s2.4}
We summarize in which points the AIBG, the almost ideal Bose gas
model, deviates from the IBG:
\begin{enumerate}
\item
The IBG wave function $\Psi_{\rm IBG}$ is multiplied by Jastrow
factors, $\Psi = F \,\Psi_{\rm IBG}$. This is a well-known
approach\cite{ch55}.
\item
By the symmetric expansion (\ref{16}) we admit a gap between the
condensed and the noncondensed particles. This modification preserves
the most basic features of the IBG, in particular the BEC mechanism and
the critical exponent $\beta =1/2$.

\item
The noncondensed single-particle states below the coherence limit
$k_{\rm coh}$ adopt the macroscopic phase of the condensate. The
leading exponent for the coherence limit $k_{\rm coh}$ is determined
from a scaling argument.
\end{enumerate}

\section{Fit to experimental data}
\label{s3}
We compare the temperature dependence of our model expression for the
superfluid density with experimental data. The model expression
contains unknown parameters; it provides a fit formula for the data. It
will turn out that this fit formula is significantly better than
comparable fit formulas.

\begin{figure}
\begin{center}
\epsfxsize=13cm
\epsfbox{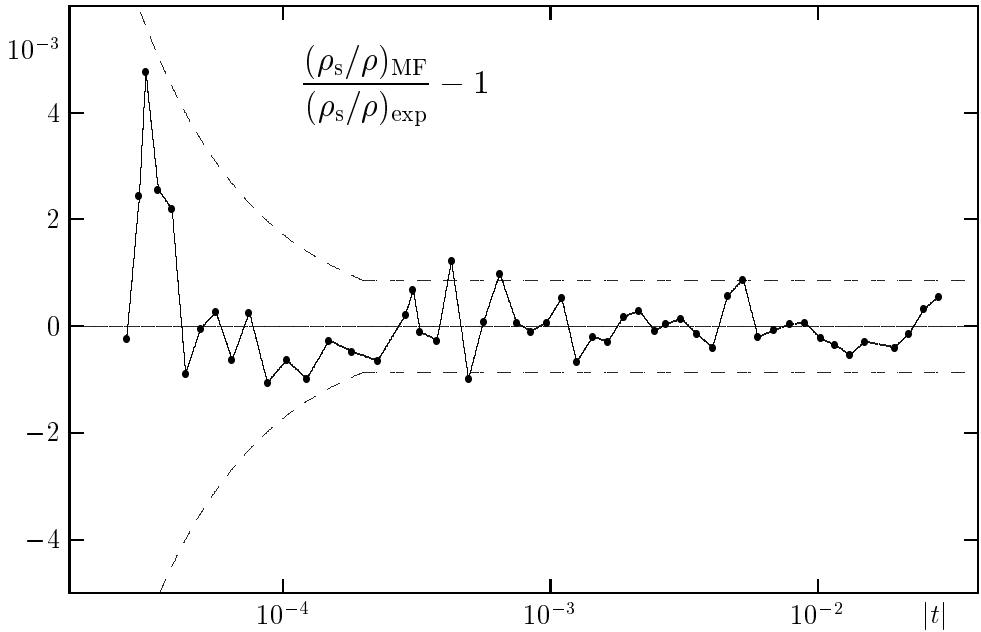}
\end{center}
\caption[]{\label{fig1}
The asymptotic model fit (MF), $(\rho_{\rm s}/\rho )_{\rm MF} =
a_1\,|t|^{2/3} + a_2\,|t| + a_3\,|t|^{4/3} $, reproduces perfectly the
experimental superfluid fraction $(\rho_{\rm s}/\rho )_{\rm exp}$. The
solid points mark the relative deviations between the fit formula and
the experimental values of Ref. \onlinecite{gr73}. As a guiding line we
draw straight lines between neighboring points. The broken line
corresponds to two standard deviations.}
\end{figure}

\subsection{Asymptotic temperature range}
\label{s3.1}
\subsubsection*{1973 data by Greywall and Ahlers}
A restriction to the first three terms in the expansion (\ref{22})
yields a three-para\-meter {\em model fit} (MF)
\begin{equation}
\label{MF}
\frac{\rho_s}{\rho} = a_1\,|t|^{2/3} +a_2\,|t| +a_3\,|t|^{4/3} 
\qquad \mbox{(MF)}
\; .
\end{equation}
Figure \ref{fig1} shows that the MF yields an excellent
reproduction of the data by Greywall and Ahlers\cite{gr73} for
saturated vapor pressure. We used all data points with temperatures
$|t|\le 0.03$.  For a minor improvement of the fit we shifted the
temperature values by $-0.5\times 10^{-7}$; this is well below the
experimental uncertainty of $\delta t = 2\times 10^{-7}$. The
parameters of the fit shown in Fig.\,\ref{fig1} are:
\begin{equation}
\label{24}
a_1 = 2.3233\,, \qquad a_2=1.0258\,, \qquad a_3=-2.0065
\; .
\end{equation}

As an alternative we consider the {\em standard fit}\/ (SF)
\begin{equation}
\label{SF}
\frac{\rho_s}{\rho} = k\,|t|^{\xi}\left(1 + D\,|t|^{\Delta}
\right) \qquad \mbox{(SF)}\,,
\end{equation}
which is used by Greywall and Ahlers\cite{gr73}, and that is
motivated by the renormalization-group theory. The fourth parameter
$\Delta$ is often\cite{gr73,go93} set equal to 1/2 because the fit is
not very sensitive to it. We will use the SF with $\Delta =0.5$ as a
three-parameter ansatz.

The fit parameters are found by minimizing the sum $\chi^2$ of the
quadratic deviations,
\begin{equation}
\label{chi2}
\chi^2  = \sum _{i=1}^{N_{\rm d}}
W\, \left[ \left( \frac{ \rho_{\rm s}}{\rho}\right)_{\!\rm fit} 
- \left( \frac{ \rho_{\rm s}}{\rho}\right)_{\!\rm exp}
\right]^{\mbox{\footnotesize 2}}
=  \sum _{i=1}^{N_{\rm d}}
\frac{1}{\sigma_{\rm rel}^2}\, 
\left[ \frac{ (\rho_{\rm s}/\rho)_{\rm fit}}
{ (\rho_{\rm s}/{\rho})_{\rm exp}} -1 \right]^{\mbox{\footnotesize 2}}
.
\end{equation}
Here $N_{\rm d}$ is the number of data points and $\sigma_{\rm rel}$ is
the relative standard deviation. The standard deviation $\sigma$ for
$(\rho_{\rm s}/\rho)_{\rm fit} - (\rho_{\rm s}/\rho)_{\rm exp}$ is
given by $W=1/\sigma^2$. The dominant experimental error is that in the
temperature. This leads to the weight\cite{gr73} $W = |t|^{2/3}/
\delta|t|^2$, where $\delta |t| = {\rm max}\,(2\times 10^{-7},
10^{-3}\,|t|)$ is the temperature uncertainty. The corresponding
$2\,\sigma_{\rm rel}$ line is shown in Fig.\,\ref{fig1}.

In the given form both MF and SF (with $\Delta =0.5$) are
three-parameter fits. We compare both fits by calculating their
$\chi^2$ ratio:
\begin{equation}
\label{27}
\frac{\chi^2_{\rm SF}}{\chi^2_{\rm MF}}\approx 8.8 \qquad 
(\mbox{data for $|t|\le 0.03$} )
\; .
\end{equation}
As seen from Fig.\,\ref{fig1} the MF reproduces the experimental data
($\chi^2/N_{\rm d}\approx 1.10$). The large ratio (\ref{27}) means that
the SF does not reproduce the data in the considered temperature range.

We remark that the SF fits the data in the considerably smaller range
$|t|\le 0.004$. This smaller range is used in Ref. \onlinecite{gr73},
presumably because it was realized that the SF does not fit the data in
the larger range. For a three-parameter fit the range $|t|\le 0.004$
appears to be rather small; we note that already a one-parameter fit
($a_1 |t|^{2/3}$) reproduces the data within 2\%{} in the relatively
large range $|t|\le 0.08$.

We considered also the data at higher pressures by Greywall and
Ahlers\cite{gr73}. Here we found ratios $\chi^2_{\rm SF}/
\chi^2_{\rm MF}$ between 1 and 2, and values of $\chi^2_{\rm MF}/N_{\rm
d}$ in the range between 3.6 and 15. This means that the MF is only
slightly better than the SF without yielding satisfactory fits.  This
is (at least partly) caused by jumps in the experimental data points.
For example, compared to a smooth fit curve (SF or MF or any reasonable
fit formula) there is a jump of more than ten standard deviations
between the data points $(|t|,\,\rho_{\rm s}/\rho) = (0.001\,439\,1,\,
0.028\,144)$ and $(0.001\,263\,1,\, 0.025\,624)$ for $P=7.27\,$bar.

\subsubsection*{1993 data by Goldner, Mulders and Ahlers}

Newer measurements of the superfluid density are reported by Goldner et
al.\cite{go93} and by Marek et al.\cite{ma88}. We consider the data by
Goldner et al. because these authors published an explicit data list.

\begin{figure}
\begin{center}
\epsfxsize=13cm
\epsfbox{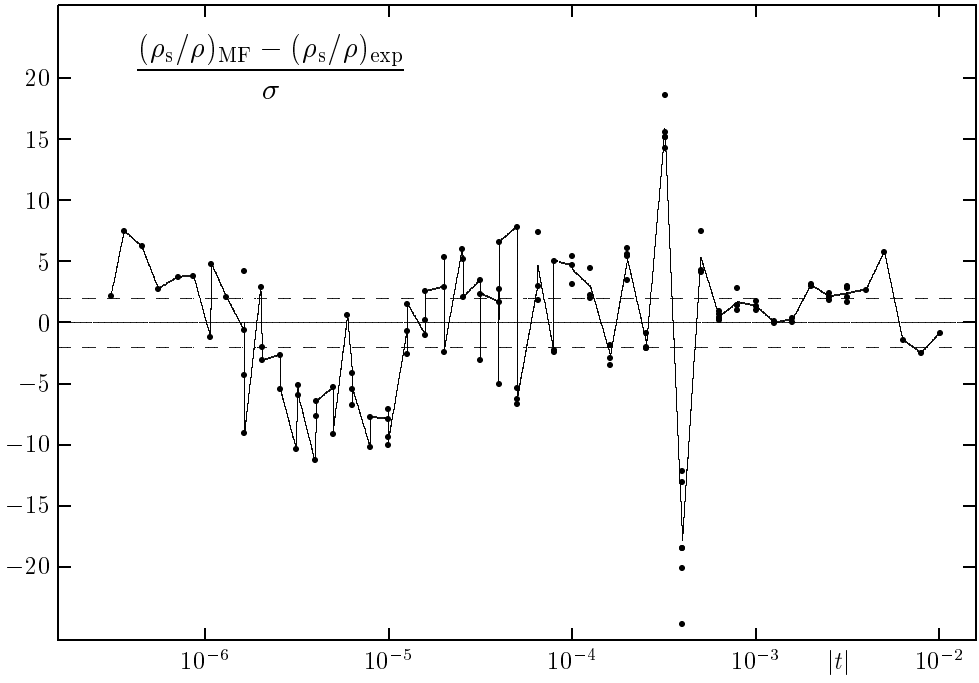}
\end{center}
\caption[]{\label{fig2}
The asymptotic model fit (MF), Eq.\,(\ref{MF}), is applied to the 1993
data of Ref. \onlinecite{go93}. The deviations between the fit and the
data are given in units of the experimental error; the broken line
corresponds to two standard deviations. For some temperatures there are
several experimental $\rho_{\rm s}$ values (each leading to a full
point); in this case the guiding line goes through the average value of
the deviation.  Sometimes the experimental temperature values are close
together (but different); this results in apparently vertical pieces of
the guiding line. From the figure it is obvious that (i) the
statistical errors are considerably larger than the assumed $\sigma$,
and (ii) the data contain several large jumps between neighboring
points.}
\end{figure}

\begin{figure}
\begin{center}
\epsfxsize=13cm
\epsfbox{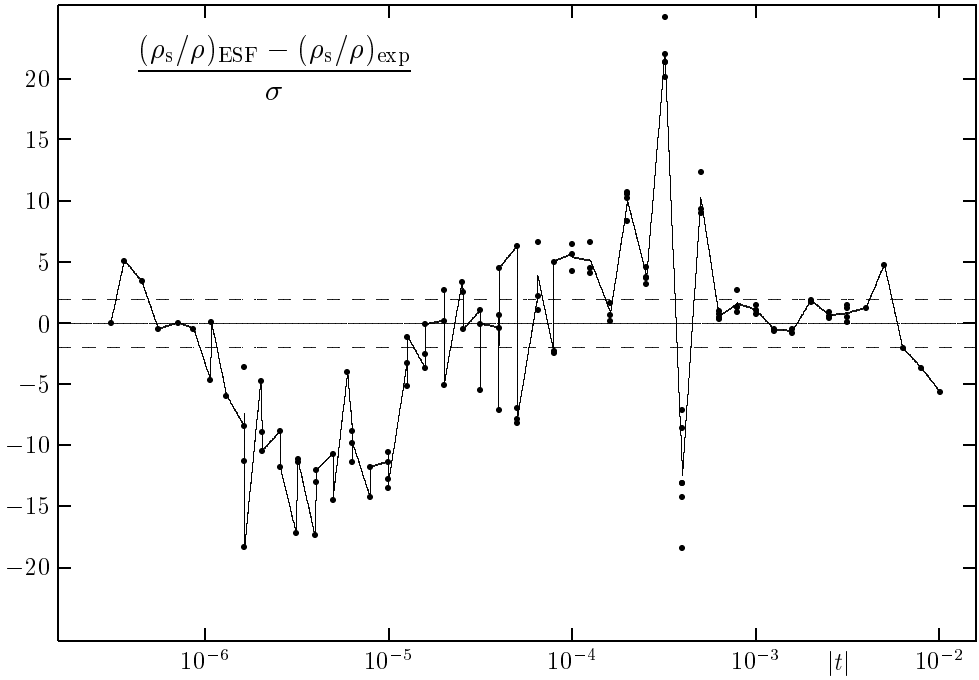}
\end{center}
\caption[]{\label{fig3} 
The extended standard fit (ESF), Eq.\,(\ref{ESF}), is applied to the
1993 data of Ref. \onlinecite{go93}. The presentation is the same as in
Fig.\,\ref{fig2}. The systematic deviations between the fit and the
data are significantly larger for this four-parameter ESF than for the
three-parameter MF (Fig.\,\ref{fig2}).}
\end{figure}

The data\cite{go93} extend to about $|t|=0.01$; all these data are used
for the fits. Fig.\,\ref{fig2} shows how the three-parameter MF
reproduces these data. We discuss this result in a number of points:
\begin{enumerate}
\item
Obviously the scatter of the data is generally larger than the
estimated error (listed as $\delta \rho_{\rm s}/\rho$ in Ref.
\onlinecite{go93}, and called $\sigma$ in Fig.\,\ref{fig2}). There are
several jumps of the size of ten standard deviations; the most dominant
jump (between the values for $|t|=0.00031910$ and $|t|=0.00039793$) is
about 30 times larger than the estimated error. This statement is
basically independent of the fit formula used (see also
Fig.\,\ref{fig3}). It is extremely unlikely that the actual superfluid
fraction contains such jumps. The different sizes of the jumps restrict
the possibility to discriminate between various fit formulas. This is
also the reason why we considered first the older 1973 data by Greywall
and Ahlers.

\item
The three-parameter SF yields a significantly larger $\chi^2$ value:
\begin{equation}
\label{28}
\frac{\chi^2_{\rm SF}}{\chi^2_{\rm MF}}\approx 2.7 \qquad 
\; .
\end{equation}
\item
Goldner et al.\cite{go93} used the following {\em extended standard
fit} (ESF)
\begin{equation}
\label{ESF}
\frac{\rho_s}{\rho} = k_0\,|t|^{\xi}\left(1 + D\,|t|^{\Delta}
\right)\left(1 + k_1\,|t|\right) \qquad \mbox{(ESF)}
\end{equation}
with $\Delta =1/2$. Using the same parameters as in Ref.
\onlinecite{go93} we obtained $\chi^2_{\rm ESF}/ \chi^2_{\rm MF}\approx
1.4$. This might appear as a small difference between MF and ESF. A
comparison between Figs.\,\ref{fig2} (MF) and \ref{fig3} (ESF) shows,
however, that the MF does a better job although it has one parameter
less. Goldner et al.\cite{go93} noted that there is a serious
discrepancy between the ESF and the data, in particular in the range
$|t| \approx 10^{-5}$ to $10^{-6}$ (their Fig.\,17). The comparison
between Fig.\,\ref{fig2} and \ref{fig3} shows that this discrepancy is
significantly smaller for our model fit. This improvement is not so
evident in the $\chi^2$ ratio because the $\chi^2$ values are on a high
level for any fit formula (due to the jumps).

\item
Looking at the scatter of the data one might tentatively assume a
standard deviation that is five times larger than the one assumed.
Drawing then a new $2\,\sigma$ line the discrepancies in the range $|t|
= 10^{-5}$ to $10^{-6}$ in Fig.\,\ref{fig2} may be judged as not very
significant.  They may, however, hint at an unexplained structure in
the temperature dependence of the superfluid fraction.
\end{enumerate}

\subsection{Extension to lower temperatures}
\label{s3.2}
We apply the model expression for the superfluid fraction in the
temperature range $1.2\,{\rm K} < T < T_\lambda$, where $|t|$ is no
longer {\em much}\/ smaller than 1. For this purpose we use the model
expressions (\ref{rho0}) and (\ref{rhoc}) for $\rho_0$ and $\rho_{\rm
coh}$, respectively, and expand $\tau$ and $x_{\rm coh}$ (rather than
$\rho_{\rm s}$ itself) into the relevant powers of $|t|$.

The expansion $\tau = a'\,|t| + b'\,t^2 +\ldots$ may be broken off
after the first term because $\tau\ne 0$ corresponds to a gap and leads
to an exponential decrease [$\propto \exp(-\tau^2)$] of the
noncondensed contribution in Eqs.\,(\ref{rho0}) and (\ref{rhoc}).
Therefore, the noncondensed contributions become rather small before
the next terms in the expansion for $\tau$ contribute significantly.

As far as the coherence limit $k_{\rm coh}$ is concerned we have no
information about the continuation of Eq.\,(\ref{21}) into an
expansion. In view of the success of Eq.\,(\ref{MF}) we will certainly
not admit exponents that would violate the form (\ref{22}). In
accordance with Eq.\,(\ref{22}) we may admit the form $x_{\rm coh} =
x_1\,|t|^{2/3} + x_2\,|t| + x_3\,|t|^{4/3} + \ldots$. This expansion
may be broken off, too, because $\rho_{\rm coh}$ of Eq.\,(\ref{rhoc})
is damped exponentially [$\propto \exp(-\tau^2)$] for increasing $|t|$.
Including the terms with the parameters $x_1$, $x_2$, and $x_3$
preserves the variability for the parameters $a_1$, $a_2$, and $a_3$ in
Eq.\,(\ref{22}).

Due to the exponential damping of the noncondensed contributions a cut
in the expansions for $\tau$ and $x_{\rm coh}$ leads much further than
a cut in the expansion for $\rho_{\rm s}/\rho$ itself. In this way we
arrive at the following {\em unified model fit} (UMF) formula:
\begin{equation}
\label{UMF}
\frac{\rho_{\rm s}}{\rho} 
 = 1 - (1+t)^{3/2}\; \frac{g_{3/2}(\tau)}{\zeta(3/2)} + \frac{4
\,(1+t)^{3/2}}{\sqrt{\pi}\, \zeta(3/2)}
\int^{x_{\rm coh}}_0 \! \frac{x^2\, dx}{\exp(x^2 +\tau^2)-1}
\qquad\mbox{(UMF)}
\end{equation}
with
\begin{equation}
\label{31}
\tau = a'\,|t|\,,\qquad 
x_{\rm coh} = \max \left( 0,\, x_1\,|t|^{2/3} + x_2\,|t| +
x_3\,|t|^{4/3}\right)
\; .
\end{equation}
As we will see, this formula provides a unified description of the
asymptotic region as well as of the less asymptotic (the ``roton'')
region.

The parameters $x_1$, $x_2$, and $x_3$ are related to the $a_1$, $a_2$,
and $a_3$ in Eq.\,(\ref{22}) and essentially fixed by the asymptotic
region.  We have restricted $x_{\rm coh}$ explicitly to non-negative
values because the expression $x_1\,|t|^{2/3} + x_2\,|t| +
x_3\,|t|^{4/3}$ might become negative for larger $|t|$ values [where,
however, the density $\rho_{\rm coh}$ tends to zero anyway because the
exponential decrease $\propto \exp(-\tau^2)$; see also
Fig.\,\ref{fig5}].

For a fit in the range $1.2\,{\rm K} < T < T_\lambda$ we combined the
data by Greywall and Ahlers\cite{gr73} for $|t|< 0.04$ and that by Clow
and Reppy\cite{cl72} (run IV) for $|t|> 0.04$.  At $|t|= 0.04$ both
data sets are compatible with each other. The systematic errors and the
deviation due to slightly different pressures (roughly 1\%{} between
saturated vapor or normal pressure) just happen to cancel each other.
Clow and Reppy\cite{cl72} remark that their ``values of $\rho_{\rm
s}/\rho$ have a scatter of about 1/2\%{}''; we interpreted this as
$\sigma_{\rm rel} = 0.005$ for our fit [i.e. for the minimization of
Eq.\,(\ref{chi2})].

\begin{figure}
\begin{center}
\epsfxsize=13cm
\epsfbox{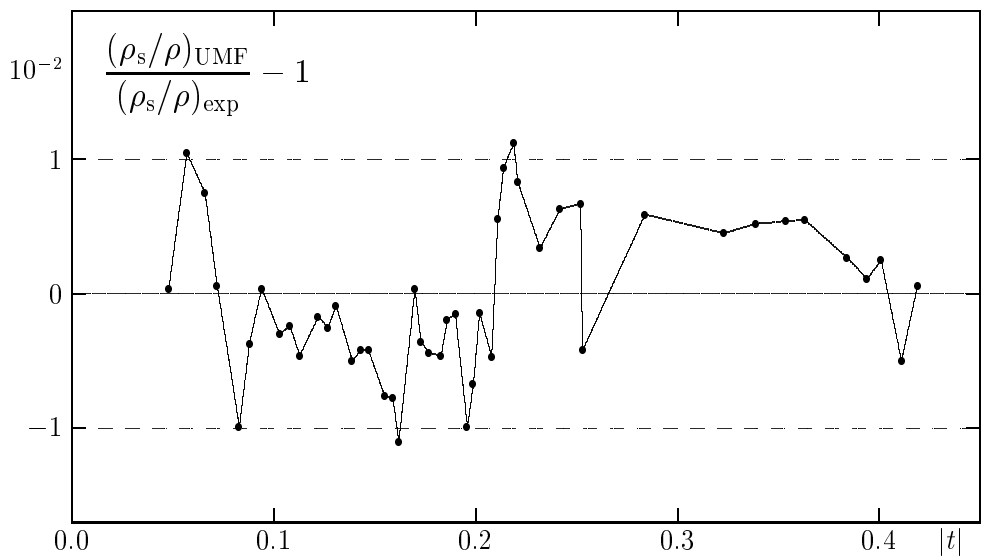}
\end{center}
\caption[]{\label{fig4} 
The unified model fit (UMF), Eq.\,(\ref{UMF}), is used to reproduce the
experimental superfluid fraction $(\rho_{\rm s}/\rho )_{\rm exp}$ in
the temperature range from $T_\lambda$ to $1.2\,$K. For $|t|\le 0.04$
we use the Greywall-Ahlers data (Ref. \onlinecite{gr73}), for $|t|>
0.04$ that of Clow and Reppy (Ref. \onlinecite{cl72}). The figure
depicts the deviations for the range $|t|> 0.04$; for $|t|< 0.04$ the
deviations are quite similar to that shown in Fig.\,\ref{fig1}. The
broken line corresponds to two standard deviations. }
\end{figure}

A fit of the combined data\cite{gr73,cl72} leads to a result that is
quite similar to Fig.\,\ref{fig1} for $|t|< 0.04$ and that is shown in
Fig.\,\ref{fig4} for $|t| > 0.04$. The fit parameters are:
\begin{equation}
\label{32}
a' = 3.0380 \,,\quad x_1 = 2.6998 \,,\quad x_2 = -0.8063 \,,\quad x_3 =
-3.9631
\qquad\mbox{(UMF)}.
\end{equation}
Alternatively we may use the parameter $f$, Eq.\,(\ref{18}), and
calculate the parameters $a_1$, $a_2$, and $a_3$ following from the
asymptotic expansion of Eq.\,(\ref{UMF}):
\begin{equation}
\label{33}
f = 5.6225 \,,\quad a_1 = 2.3323 \,,\quad a_2 = 0.8035 \,,\quad a_3 =
-0.4704 \;.
\end{equation}
If an expansion is broken off as in Eq.\,(\ref{MF}) the last term tries
effectively to simulate the missing terms. Since the UMF supplies
higher-order terms it is not surprising that the last coefficients in
Eqs.\,(\ref{24}) and (\ref{33}) are quite different.

Alternatively we used the data by Tam and Ahlers\cite{ta87} that
extend, however, only down to 1.5\,K. This yields similar parameter
values.

The standard fit for temperatures above $1\,{\rm K}$ but excluding the
asymptotic region is the two-parameter {\em roton fit}\/ (RF),
\begin{equation}
\label{RF}
\frac{\rho_{\rm s}}{\rho} =  \frac{A}{\sqrt{T}}\;
\exp\left( - \frac{\Delta}{k_{\rm B} T} \right) \qquad \mbox{(RF)}
\; .
\end{equation}
Using the data of Ref. \onlinecite{cl72} (run IV) we obtain
\begin{equation}
\label{35}
\frac{\chi^2_{\rm RF}}{\chi^2_{\rm UMF}}\approx 4 \qquad 
\mbox{for $1.2\,{\rm K} < T < 2.07\,{\rm K}$}
\, .
\end{equation}
This ratio is reduced to 2 if we restrict the temperature by $T < 2\,
{\rm K}$. These ratios imply that the unified model expression is quite
good for intermediate temperatures, too.

The RF is based on Landau's quasiparticle model that cannot be extended
to $T_\lambda$ without loosing its physical basis. The standard
description for the range $1.2\,{\rm K} < T < T_\lambda$ would be a
combination of the SF (\ref{SF}) and the RF (\ref{RF}). In contrast to
this, our model provides a unified fit (\ref{UMF}) in this range.
Although containing one parameter less (than the combination of SF and
RF) this unified fit is superior to the standard description.

As already mentioned, the expansion (\ref{16}) implies a gap between
the condensed and noncondensed particles. This gap appears to be
essential for the reproduction of the data in the intermediate range
$T\gtrsim 1\,{\rm K}$. This gap should in some way be related to the
roton gap $\Delta$. This relation cannot be expected to be simple and
obvious because one gap belongs to a model (Landau) for $T\ll
T_\lambda$ and the other to a model (AIBG) for $T\sim T_\lambda$. We
note that our gap vanishes for $T\to T_\lambda$, and that the roton
concept becomes less sharp for increasing temperature (for $T = 1\,{\rm
K}$ the widths of roton states are already comparable to their
energies).

For $T\ll T_\lambda$ Landau's quasiparticle model is, of course, the
right model. The model fit (\ref{UMF}) yields still reasonable values
for $\rho_{\rm s}/\rho$ but it must fail in the quantitative
reproduction of $1-\rho_{\rm s}/\rho$ because the phonons are not
described by the wave function (\ref{Psic}).

\section{Condensate density}
\label{s4}

\begin{figure}
\begin{center}
\epsfbox{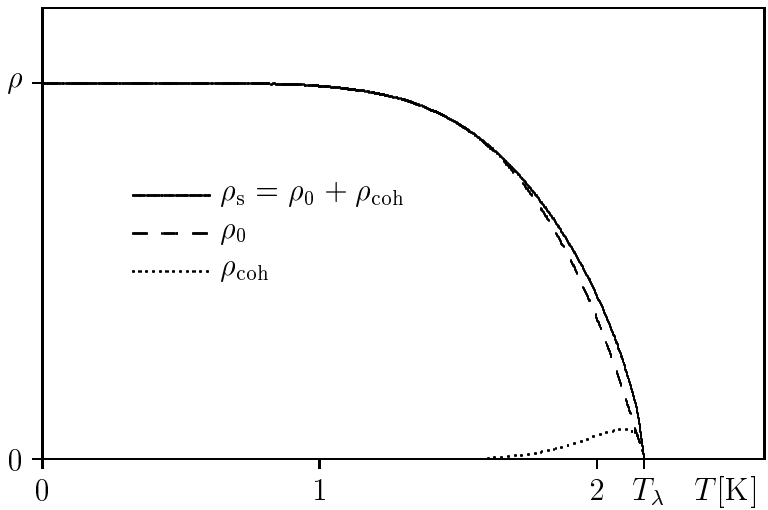}
\end{center}
\caption[]{\label{fig5} 
Decomposition of the superfluid density $\rho_{\rm s}$ into the (model)
condensate density $\rho_0$ and the coherently comoving density
$\rho_{\rm coh}$ as a function of the temperature $T$.  For $T\to
T_\lambda$ the comoving density $\rho_{\rm coh}\propto |t|^{2/3}$ is
the dominant contribution to the superfluid density. The full line
represents the model expression for $\rho_{\rm s}$. In the given scale,
this curve coincides with the experimental superfluid density for all
temperature values.}
\end{figure}

The unified model fit, Eq.\,(\ref{UMF}) with Eq.\,(\ref{32}), defines
the decomposition of the superfluid density into the condensate density
and the coherently comoving density. The temperature dependence of this
decomposition is displayed in Fig.\,\ref{fig5}. In this section we
discuss in particular the temperature dependence of the condensate
density.

The contribution of $\rho_{\rm coh}$ is decisive near $T_\lambda$ but
negligible for lower temperatures. The comoving density $\rho_{\rm
coh}$ carries some entropy because it does not correspond to a single
quantum state. This entropy content is quite small because it is due to
the lowest single-particle states with $\langle n_{\bf k}\rangle \gg
1$. It is below the present experimental limits but should be
detectable; for these points we refer to Refs. \onlinecite{fl91} and
\onlinecite{sc94}.

As shown in Sec.\,\ref{s3.2}, the expression $\tau = a'\,|t|$ works
quite well for fitting the data down to about 1.2\,{\rm K}.  Inserting
$\tau = a'\,|t|$ in Eq.\,(\ref{rho0}) yields
\begin{equation}
\label{36}
\frac{\rho_{\rm 0}}{\rho} 
 = 1 - (1+t)^{3/2}\; \frac{g_{3/2}(a'\,|t|)}{\zeta(3/2)}
\; .
\end{equation}
Using $a'$ of Eq.\,(\ref{32}), this temperature dependence is shown by
the dashed line in Fig.\,\ref{fig5}.

The asymptotic expansion of Eq.\,(\ref{36}) reads $\rho_{\rm 0}/\rho
\sim f|t|$, Eq.\,(\ref{17}), where
\begin{equation}
\label{37}
f = \frac{3}{2} + \frac{2\sqrt{\pi }\,a'}{\zeta (3/2)} \approx 5.6
\; .
\end{equation}
The numerical value is taken from Eq.\,(\ref{33}). We found that
\begin{equation}
\label{38}
\frac{\rho_{\rm 0}}{\rho} 
\approx  1 - \Bigg( \frac{T}{T_\lambda}\Bigg)^{\!\mbox{\footnotesize $f$}}
\; .
\end{equation}
may be used as an approximation for Eq.\,(\ref{36}). The maximum
relative difference between Eqs.\,(\ref{38}) and (\ref{36}) is about
2\%{}. For $T\to T_\lambda$ both expressions, (\ref{38}) and
(\ref{36}), yield $\rho_0/\rho \sim f|t|$.

The right-hand side of Eq.\,(\ref{38}) is an old fit formula for the
superfluid fraction $\rho_{\rm s}/\rho$ (for example, Fig.\,27 of Ref.
\onlinecite{lo54}). In the framework of our model, this historic fit
formula may be interpreted as the approximation $\rho_{\rm s}\approx
\rho_0$. The obvious shortcomings of Eq.\,(\ref{38}) as an
approximation for $\rho_{\rm s}/\rho$ are the following: (i) The
neglect $\rho_{\rm coh}$ leads to a qualitatively wrong asymptotic
behavior (difference between the full and the dashed line in
Fig.\,\ref{fig5}). (ii) The step from Eq.\,(\ref{36}) to
Eq.\,(\ref{38}) as well as the use of $\tau = a'|t|$ make the
expression an approximate one already for $\rho_0/\rho$. (iii) For
small temperatures $1-\rho_{\rm s}/\rho = (T/T_\lambda)^f$ is
quantitatively wrong (because the phonons have not been taken into
account).

We consider once more the exact condensate fraction $\rho^{\rm
exact}_0/\rho$ introduced in Sec.\,\ref{s2.2}. We denote its value at
$T=0$ by $n_{\rm c}$.  Assuming that the depletion of the condensate
(from 1 to $n_{\rm c}\approx 0.1$) is temperature independent we obtain
\begin{equation}
\label{39}
\frac{\rho^{\rm exact}_0}{\rho} \approx  n_{\rm c}\; \frac{\rho_0}{\rho}
\approx n_{\rm c} \Bigg( 1- \frac{T^f}{T^{\, f}_\lambda}\Bigg)
\; .
\end{equation}
as an approximate expression for the temperature dependence of the
exact condensate fraction. The experimental temperature dependence is
given in Fig.\,2 of Snow et al.\cite{sn92}. Within the relatively large
experimental uncertainties the expression (\ref{39}) agrees with the
data.

\section{Effective Ginzburg-Landau model}
\label{s5}

In our approach, the macroscopic wave function (\ref{psi}),
\begin{equation}
\label{40}
\psi ({\bf r}) = \sqrt{\frac{n_0}{V}}\;
\exp\big[\,{\rm i}\,\Phi ({\bf r})\big]
= \sqrt{\rho_0}\,
\exp\big[\,{\rm i}\,\Phi ({\bf r})\big]  \; ,
\end{equation}
plays the role of the order parameter. We investigate the free energy
as a function of this order parameter.

\subsection{Coherence limit}
\label{s5.1}

We start by presenting a qualitative argument for the existence and the
meaning of the coherence limit $k_{\rm coh}$.

The macroscopic wave function $\psi$ may contain equilibrium and
nonequilibrium excitations. A superfluid motion with ${\bf u}_{\rm s} =
(\hbar/m) \nabla\Phi$ is a nonequilibrium excitation. At finite
temperatures, there are thermal fluctuations of the order parameter,
i.e., equilibrium excitations. We consider the average momentum of these
fluctuations,
\begin{equation}
\label{41}
k_{\rm fluct} = \overline{ | \nabla \Phi | }
\; .
\end{equation}
The bar denotes the statistical average. The momentum $k_{\rm fluct}$
will be a function of the temperature. It is related to the correlation
length $\xi \approx 1/k_{\rm fluct}$.

The single-particle states are described by real functions
$\varphi_{\bf k}$ in Eq.\,(\ref{Psi}). We consider the possibility of
phase fluctuations for a low-lying state with $n_{\bf k} \gg 1$, too.
After the replacement $\varphi_{\bf k} \longrightarrow \varphi_{\bf
k}\exp ({\rm i}\,\Phi_{\bf k})$ in Eq.\,(\ref{Psi}) these fluctuations
may be described by the fields $\Phi_{\bf k}({\bf r})$.

Let us first assume that the additional phases vanish, $\Phi_{\bf
k}=0$. In this case, the average kinetic energy $\hbar^2 k_{\rm
fluct}^{\,2}/2 m$ of a condensed particle would exceed that of a
noncondensed particle with $k<k_{\rm fluct}$. The energy sequence of
the single-particle states is, however, a prerequisite of the BEC; the
condensate must be formed by the particles with the lowest energy. In
order to preserve the energy sequence of the low-lying states we
require the phase ordering
\begin{equation}
\label{42}
\Phi_{\bf k}({\bf r}) = \Phi({\bf r}) \quad \mbox{for}\quad
k \le k_{\rm fluct}
\; .
\end{equation}
This argument does not apply to the states with higher momenta. By this
qualitative argument we obtain the many-body wave function (\ref{Psic})
with
\begin{equation}
\label{43}
k_{\rm coh} = k_{\rm fluct}
\; .
\end{equation}

\subsection{Free energy} 
\label{s5.2}

The statistical expectation value $\rho_0 \sim |t|$ can be obtained by
minimizing the common Landau energy $F_{\rm L}/V = R\,t\,|\psi|^2
+U\,|\psi|^4$ (with regular coefficients $R$ and $U$). The fluctuation
term $F_{\rm fluct}/V = (\hbar^2/2m)\,|\nabla\psi|^2$ equals the
kinetic energy density $\rho_0 \,{\bf u}^2/2$  of the condensate only;
here ${\bf u} = (\hbar/m)\nabla \Phi$. The phase coherence assumed in
Eq.\,(\ref{Psic}) implies that $\rho_0 {\bf u}^2$ must be replaced by
$\rho_{\rm s} {\bf u}^2$. This leads to the following {\em effective}\/
Ginzburg-Landau ansatz
\begin{equation}
\label{GL}
\frac{F_{\rm GL}}{V} = \frac{F_{\rm fluct}  + F_{\rm L}}{V} = 
\frac{\hbar^2}{2 m}\,\frac{\rho_s}{\rho_0}\;
|\nabla\psi|^2 +R\,t\,|\psi|^2 +U\,|\psi|^4
\; .
\end{equation}
Assuming that the leading exponent of $x_{\rm coh}$ is not greater than
1, Eq.\,(\ref{20}) yields
\begin{equation}
\label{45}
\rho_{\rm coh} \sim x_{\rm coh} \; .
\end{equation}
The equilibrium fluctuation term becomes then
\begin{equation}
\label{46}
F_{\rm fluct} \propto \rho_s \, k_{\rm fluct}^{\,2} = (\rho_0 +
\rho_{\rm coh}) \, k_{\rm coh}^{\,2}
\sim \rho_0 \, x_{\rm coh}^{\,2} + x_{\rm coh}^{\,3}
\; .
\end{equation}
The asymptotic form of the Landau part of the free energy behaves like
\begin{equation}
\label{47}
F_{\rm L} \propto  R\,t\,\rho_0 +U\,\rho_0^{\,2} \sim  t^2
\; .
\end{equation}
We require now scaling invariance. This means that $F_{\rm fluct}$ must
have the same leading $|t|$ dependence as $F_{\rm L}$, i.e., $\rho_0 \,
x_{\rm coh}^{\,2} + x_{\rm coh}^{\,3} \sim t^2$.  From $\rho_0 \,x_{\rm
coh}^{\,2} \sim t^2$ we would obtain $x_{\rm coh} \sim |t|^{1/2}$ and
$x_{\rm coh}^{\,3}\sim |t|^{3/2}$ in contradiction to the scaling
assumption. Therefore, scaling requires $x_{\rm coh}^{\,3}\sim |t|^2$
or
\begin{equation}
\label{48}
k_{\rm coh}\sim |t|^{2/3}
\; .
\end{equation}
This implies $\rho_s\sim\rho_{\rm coh}\sim |t|^{2/3}$ for the
superfluid density and $\xi \approx 1/k_{\rm fluct}\sim |t|^{-2/3}$ for
the correlation length.

The mass coefficient $\rho_s/\rho_0\sim |t|^{-1/3}$ in Eq.\,(\ref{GL})
is singular. Ginzburg and Sobyanin\cite{gi82} have introduced a
comparable effective Ginzburg-Landau model with nonanalytic
coefficients, too. In Ref. \onlinecite{gi82} the nonanalytic
coefficients (like $|t|^{4/3}$ for the $|\psi|^2$ term) are
phenomenologically introduced in order to reproduce the right critical
exponents.

The divergent mass coefficient $\rho_s/\rho_0\sim |t|^{-1/3}$ damps the
critical fluctuations such that Eq.\,(\ref{GL}) becomes scaling
invariant.  In this sense, the model (\ref{GL}) has properties similar
to the common Ginzburg-Landau ansatz in $d=4$ dimensions. This means
that Eq.\,(\ref{GL}) might be used down to $|t|=0$ and that the
critical exponent of $\rho_s$ might be indeed exactly $2/3$.  This
possibility is supported by the excellent fit obtained for
Eq.\,(\ref{MF}).

\subsection{Further scaling restrictions}
\label{s5.3}

The equilibrium Landau free energy contains integer powers of $t$ only:
\begin{equation}
\label{49}
F_{\rm L}  \sim ...\,t^2 + ... \,t^3 +...\, t^4 +...
\; .
\end{equation}
The asymptotic form of the superfluid density (\ref{22}) is compatible
with the expansion $x_{\rm coh} \propto x_1\,|t|^{2/3} + x_2\,|t| +
x_3\,|t|^{4/3} +\ldots$. In the fluctuation term this expansion will,
however, in general lead to noninteger exponents:
\begin{equation}
\label{50}
F_{\rm fluct}  \propto \rho_{\rm s}\,k_{\rm coh}^{\,2} = ( \rho_0 +
\rho_{\rm coh})\,k_{\rm coh}^{\,2}
\sim ...\, t^2 + ... \, |t|^{7/3}+ ...\, |t|^{8/3}+ ...\, t^{3} +...
\; .
\end{equation}
Scaling for Eq.\,(\ref{GL}) implies also that the amplitudes of the
nonanalytic terms vanish. This condition yields relations between the
expansion parameters $a'$, $b'$,\,\ldots{} and $x_1$, $x_2$,
$x_3$,\,\ldots{} that may also be expressed by the coefficients $a_i$
in Eq.\,(\ref{22}). The condition of a vanishing amplitude of the
$|t|^{7/3}$ term can be evaluated straightforwardly and yields
\begin{equation}
\label{51}
x_2 = - \frac{\sqrt{\pi}\,\zeta(3/2)}{8} \approx - 0.58
\quad \mbox{ or }\quad a_2=1
\; .
\end{equation}
These theoretical values compare well with the fitted values given in
Eqs.\,(\ref{24}) or (\ref{33}).

The condition of a vanishing amplitude of the $|t|^{8/3}$ term yields
\begin{equation}
\label{52}
x_3 = \frac{\pi\,[\zeta(3/2)]^2}{64\, x_1}-\frac{a'^2}{3\, x_1}
\end{equation}
and a corresponding expression for $a_3$. These relations are not
fulfilled by the parameter values found in the fits. The reason is
probably the following: The expansions (\ref{22}) and (\ref{31}) are
cut after the $|t|^{4/3}$ term. In a fit it is then in particular the
last term that tries to simulate the neglected terms.

\section{Concluding remarks}
\label{s6}

We have modified the IBG in such a way that it might be applied to
liquid helium. We summarize the novel views and main results of our
approach.
\begin{enumerate}
\item
The IBG value $\beta =1/2$ for the critical exponent of the condensate
should be taken seriously. It is not subject to renormalization because
it results from a calculation that already includes a summation over
arbitrarily large lengths, and it is essential for the BEC mechanism.
\item
The model condensate contributes fully to the superfluid density; it is
not depleted by the Jastrow factors.
\item
In order to reproduce the critical exponent $\nu \approx 1/3$ of the
superfluid density we have assumed that noncondensed particles below a
certain momentum $k_{\rm coh}$ move coherently with the condensate. The
coherence limit $k_{\rm coh}$ has been made plausible in
Sec.\,\ref{s5.1}.

The contribution of noncondensed particles to the superfluid density
offers a solution of the so-called macroscopic problem\cite{uh73} of
liquid helium. This problem reads as follows: If the superfluid density
corresponds to single quantum state ($\rho_{\rm s}\propto |\psi|^2$)
then the approach to an equilibrium state [with $\rho_{\rm s} =
\rho_{\rm s}(T)$] cannot be understood.

\item
We have derived a fit formula for the temperature dependence of the
superfluid density. This fit formula reproduces the data significantly
better than comparable expressions. This feature as well as qualitative
scaling arguments suggest that the critical exponent $\nu$ of the
superfluid density might be exactly equal to 2/3.

\item 
The temperature dependence of the decomposition of superfluid density
into the model condensate density and the coherently comoving density
is given. A simple formula for the temperature dependence of the
depleted condensate density is presented.
\end{enumerate}


\end{document}